\documentclass[ALICE,manyauthors]{cernphprep}
\usepackage{hyperref}
\usepackage{lineno}
\usepackage{xcolor}
\usepackage{euscript}
\usepackage{cite}
\usepackage[T1]{fontenc}
\usepackage[Symbolsmallscale]{upgreek}
\usepackage{multirow}
\newcommand{\pt}           {\ensuremath{p_{\rm T}}\xspace}

\newcommand{\rs}           {\ensuremath{r^*}\xspace}
\newcommand{\rsv}           {\ensuremath{\vec{r}^*}\xspace}

\newcommand{\ks}           {\ensuremath{k^*}\xspace}
\newcommand{\kst}           {\ensuremath{k^*_\mathrm{true}}\xspace}

\newcommand{\ksv}           {\ensuremath{\vec{k}^*}\xspace}

\newcommand{\pL}           {\ensuremath{\mbox{p}\Lambda}\xspace}
\newcommand{\decpLpL}           {\ensuremath{\mathrm{p}\Lambda}\xspace}
\newcommand{\decpLpSo}           {\ensuremath{\mathrm{p}(\Sigma^0)}\xspace}
\newcommand{\decpLpXi}           {\ensuremath{\mathrm{p}(\Xi)}\xspace}

\newcommand{\decpLflat}           {\ensuremath{\mathrm{ff}}\xspace}
\newcommand{\decpfL}           {\ensuremath{\mathrm{p}\tilde{\Lambda}}\xspace}
\newcommand{\decfpL}           {\ensuremath{\tilde{\mathrm{p}}\Lambda}\xspace}

\newcommand{\pSo}          {\ensuremath{\mbox{p\So}}\xspace}
\newcommand{\pXi}          {\ensuremath{\mbox{p}\Xi}\xspace}
\newcommand{\pXio}          {\ensuremath{\mbox{p}\Xi^0}\xspace}
\newcommand{\pXim}          {\ensuremath{\mbox{p\X}}\xspace}

\newcommand{\nineH}        {$\sqrt{s}~=~0.9$~Te\kern-.1emV\xspace}
\newcommand{\seven}        {$\sqrt{s}~=~7$~Te\kern-.1emV\xspace}
\newcommand{\onethree}        {$\sqrt{s}~=~13$~Te\kern-.1emV\xspace}
\newcommand{\twoH}         {$\sqrt{s}~=~0.2$~Te\kern-.1emV\xspace}
\newcommand{\twosevensix}  {$\sqrt{s}~=~2.76$~Te\kern-.1emV\xspace}
\newcommand{\five}         {$\sqrt{s}~=~5.02$~Te\kern-.1emV\xspace}
\newcommand{\twosevensixnn}{$\sqrt{s_{\mathrm{NN}}}~=~2.76$~Te\kern-.1emV\xspace}
\newcommand{\fivenn}       {$\sqrt{s_{\mathrm{NN}}}~=~5.02$~Te\kern-.1emV\xspace}

\newcommand{\MeV}  {\ensuremath{\text{Me\kern-.1emV}}\xspace}
\newcommand{\MeVc}  {\ensuremath{\text{Me\kern-.1emV/}c}\xspace}
\newcommand{\MeVcc}  {\ensuremath{\text{Me\kern-.2emV/}c^2}\xspace}
\newcommand{\GeV}  {\ensuremath{\text{Ge\kern-.1emV}}\xspace}
\newcommand{\GeVc}  {\ensuremath{\text{Ge\kern-.1emV/}c}\xspace}
\newcommand{\GeVcc}  {\ensuremath{\text{Ge\kern-.2emV/}c^2}\xspace}
\newcommand{\TeV}  {\ensuremath{\text{Te\kern-.1emV}}\xspace}

\newcommand{\lmb}          {\ensuremath{\Lambda}\xspace}
\newcommand{\almb}         {\ensuremath{\overline{\Lambda}}\xspace}
\newcommand{\So}           {\ensuremath{\Sigma^{0}}\xspace}

\newcommand{\X}            {\ensuremath{\Xi^-}\xspace}

\newcommand{\SN}{\ensuremath{\rm N \Sigma}\xspace}
\newcommand{\LN}{\ensuremath{\rm N \Lambda}\xspace}
\newcommand{\SNLN}{\ensuremath{\rm N \Sigma\leftrightarrow N \Lambda}\xspace}

\newcommand{\Chieft}           {\ensuremath{\rm \chi EFT}\xspace}

\begin{document}%

%%%%%%%%%%%%%%%  Title page %%%%%%%%%%%%%%%%%%%%%%%%
\begin{titlepage}
\PHyear{2021}
\PHnumber{51}      % required, will be obtained from PH
\PHdate{30 March}  % required, will be obtained from PH
%

%%% Put your own title + short title here:
\title{Exploring the N$\Lambda$--N$\Sigma$ coupled system with high precision correlation techniques at the LHC}
\ShortTitle{Exploring the N$\Lambda$--N$\Sigma$  system with high precision correlation techniques}   % appears on right page headers

%%% Do not change the next lines
\Collaboration{ALICE Collaboration\thanks{See Appendix~\ref{app:collab} for the list of collaboration members}}
\ShortAuthor{ALICE Collaboration} % appears on left page headers, do not change

\begin{abstract}

The interaction of $\Lambda$ and $\Sigma$ hyperons (Y) with nucleons (N) is strongly influenced by the coupled-channel dynamics. Due to the small mass difference of the \LN and \SN systems, the sizeable coupling strength of the \SNLN processes constitutes a crucial element in the determination of the N$\Lambda$ interaction. In this letter we present the most precise measurements on the interaction of p$\Lambda$ pairs, from zero relative momentum up to the opening of the \SN channel. The correlation function in the relative momentum space for $\mathrm{p}\lmb\oplus\overline{\mathrm{p}}\almb$ pairs measured in high-multiplicity triggered pp collisions at \onethree at the LHC is reported. The opening of the inelastic N$\Sigma$ channels is visible in the extracted correlation function as a cusp-like structure occurring at relative momentum {$\ks=289~\MeVc$}. This represents the first direct experimental observation of the \SNLN coupled channel in the p$\Lambda$ system. The correlation function is compared with recent chiral effective field theory calculations, based on different strengths of the \SNLN transition potential. A weaker coupling, as possibly supported by the present measurement, would require a more repulsive three-body NN$\Lambda$ interaction for a proper description of the $\Lambda$ in-medium properties, which has implications on the nuclear equation of state and for the presence of hyperons inside neutron stars.
\end{abstract}
\end{titlepage}

\setcounter{page}{2}
%\our paper goes here:\\~\\

\section{Introduction}\label{sec:Intro}

The proton--Lambda (p$\Lambda$) system is one of the best-known examples in hadron physics where the role of coupled-channel dynamics is crucial for the understanding of the two-body and three-body interaction, both in vacuum and at finite nuclear densities~\cite{Polinder:2006zh,Haidenbauer:2013oca,Haidenbauer:2019boi,Gerstung:2020ktv}. The coupling between the nucleon--Sigma (\SN) and \LN systems arises from these pairs having the same strangeness content and a small mass difference, and it is responsible for the dominant attractive p$\Lambda$ interaction in the spin-triplet state of coupled-channel potentials~\cite{Haidenbauer:2019boi,Petschauer:2020urh,Nagels:2015lfa}.

The attractive nature of the interaction between a proton and a \lmb was established from measurements of binding energies of light \lmb-hypernuclei~\cite{Hashimoto:2006aw,Gal:2016boi} and scattering experiments at low energies~\cite{SechiZorn:1969hk,Alexander:1969cx,Eisele:1971mk}. 
However, the available scattering cross sections are characterised 
by large uncertainties. Moreover, they are limited to hyperon 
momenta above {$p_\mathrm{lab}\sim$100~\MeVc}. Thus, a reliable
determination of standard quantities like scattering lengths,
which provide a simple quantitative measure for the strength 
of an interaction, is practically impossible. Furthermore, 
in the region $p_\mathrm{lab} \approx 640$~\MeVc, where the n$\Sigma^+$ and p$\Sigma^0$ channels open, the momentum resolution 
of the existing data is poor~\cite{Kadyk:1971,Hauptman:1977}.
Calculations based on N$\Lambda$-N$\Sigma$ coupled-channel potentials
\cite{Haidenbauer:2013oca,Haidenbauer:2019boi,Nagels:2015lfa} 
predict a narrow but sizeable enhancement of the p$\Lambda$ 
cross section
in that region which reflects the strength of the channel coupling
and also that of the N$\Sigma$ interaction. However, because of the
poor resolution of the mentioned scattering data, the presence of 
such a structure could not be confirmed. New p$\Lambda$ data that
became available recently~\cite{CLAS:2021gur} cover only energies
well above the N$\Sigma$ threshold. 
Experimental observations of a cusp-like structure at the \SN threshold stem only from studies of the p$\Lambda$ invariant mass (IM) spectrum in strangeness exchange processes such as $\mathrm{K^-}\mathrm{d}\rightarrow\uppi^-\mathrm{p}\Lambda$~\cite{Tan:1969jq,Braun:1977ma} and more recently from measurements of the reaction $\mathrm{pp}\rightarrow\mathrm{K^+}\mathrm{p} \Lambda$~\cite{ElSamad:2012kg,Munzer:2017hbl}.

It is known that the strength of the \SNLN conversion is relevant for the behaviour of $\Lambda$ hyperons in infinite nuclear matter~\cite{Nogami:1970,Bodmer:1971,Dabrowski:1973}. This has been emphasised in a recent study of the YN interaction based on chiral effective field theory ($\chi$EFT)~\cite{Haidenbauer:2019boi}. Specifically, this work discussed the interplay between the \SNLN conversion, the in-medium properties of the $\Lambda$ and the role played by three-body forces. The abundant data on hypernuclei allowed the determination of the average attraction ($-30$~\MeV) experienced by a $\Lambda$ hyperon within symmetric nuclear matter at the nuclear saturation density~\cite{Tolos:2020aln}. However, the interaction of hyperons with the surrounding nucleons at larger baryonic densities is not known empirically. 
The outcome of pertinent calculations depends on the employed \LN and NN$\Lambda$ interactions in vacuum. These contributions are directly correlated to the \SNLN conversion, as the parameters driving the coupling strength in the theory can be tuned differently while still reproducing the existing experiments~\cite{Haidenbauer:2019boi}. For example, compared to the original version of the next-to-leading order (NLO) $\chi$EFT (NLO13)~\cite{Haidenbauer:2013oca}, the revisited version (NLO19)~\cite{Haidenbauer:2019boi} involves a weaker \SNLN transition potential. However, it leads to practically identical results for \LN two-body scattering, but to an enhanced attractive behaviour in the medium. This points to a stronger repulsive three-body force needed within the latter realisation. 
The interplay between the \LN and NN$\Lambda$ interaction is relevant to the debated presence of $\Lambda$ hyperons inside the core of neutron stars (NS)~\cite{Djapo:2008au,Tolos:2020aln,Logoteta:2019}. The hyperon puzzle originates from the contraposition between the energetically favoured production of hyperons in the interior of NS~\cite{Vidana:2000ew} and the subsequent softening of the corresponding equation of state (EoS). The latter does not support the existence of the heaviest observed NS of up to $2.2$ solar masses~\cite{NS1Demorest:2010bx,NS2Antoniadis:2013pzd,NS3Cromartie:2019kug}. Applications of the NLO19 $\chi$EFT potentials in calculations of the EoS~\cite{Gerstung:2020ktv} demonstrated that a repulsive genuine NN$\Lambda$ interaction suppresses the appearance of $\Lambda$ hyperons inside NS, giving a more quantitative reference for the solution of the hyperon puzzle. Thus new experimental data of high precision providing constraints on the \SNLN dynamics are needed.

Recent studies of two-particle correlations in pp, p--Pb and Pb--Pb collisions have been successful in studying the final-state interaction (FSI) and in delivering high precision data on particle pairs of limited accessibility using traditional experimental techniques~\cite{ALICE:Run1,Acharya:2020dfb,ALICE:LL,ALICE:pSig0,ALICE:pXi,ALICE:pOmega,ALICE:pK,ALICE:KK_pp,ALICE:KK_HI,ALICE:LK_HI,ALICE:BantiB_HI}. 
Performing such measurements in small collision systems results in a stronger sensitivity of the experimental correlation to the coupled-channel dynamics, as recently proven by means of $\mathrm{p}\mathrm{K}^-$ correlations~\cite{ALICE:pK,Haidenbauer:2018jvl,Kamiya:2019uiw}.
In this letter we present the combined measurement of $\mathrm{p}\lmb$ and $\overline{\mathrm{p}}\almb$ pairs in pp collisions with a high-multiplicity (HM) trigger at $\sqrt{s}=$13~\TeV~\cite{ALICE,ALICEperf}.

\section{Data analysis}\label{sec:Data}

The relevant observable in this analysis is the two-particle correlation function $C(\ks)$. This is related to an effective particle emission source $S(\rs)$ and to the wave function $\Psi(\ksv,\rsv)$ of the particle pair, by means of the relation $C(\ks)=\int S(\rs)|\Psi(\ksv,\rsv)|^2d^3\rs$~\cite{Lisa:2005dd}, where the relative distance \rs and relative momentum $q^*=2\ks$ are evaluated in the pair rest frame. The experimental correlation is defined as 
\begin{equation}\label{eq:Cks}
    C(\ks)=\EuScript{N}\cdot N(\ks)/M(\ks),
\end{equation}
where $N(\ks)$ is the distribution of pairs where both reconstructed particles are measured in the same event, $M(\ks)$ is the reference distribution of uncorrelated pairs sampled from different (mixed) events and $\EuScript{N}$ is a normalisation factor. The uncorrelated sample in the denominator, $M$(\ks), is obtained by combining particles from one event with particles from a set of other events. The two events are required to have comparable number of charged particles at midrapidity and a similar primary vertex coordinate $V_z$ along the beam axis ($z$).

The ALICE experiment excels in correlation studies thanks to its good tracking and particle identification (PID)~\cite{ALICE,ALICEperf}. These capabilities are related to the three subdetectors, the inner tracking system (ITS)~\cite{ALICEITS}, the time projection chamber (TPC)~\cite{ALICETPC} and the time-of-flight detector (TOF)~\cite{ALICETOF}. The event trigger is based on the measured amplitude in the V0 detector system, consisting of two arrays of plastic scintillators located at forward ($2.8<\eta<5.1$) and backward ($-3.7<\eta<-1.7$) pseudorapidities~\cite{Abbas:2013taa}. The selected HM events correspond to 0.17\% of all events with at least one measured charged particle within $|\eta|<1$ (INEL$>0$). This condition results in an average of 30 charged particles in the range $|\eta|<0.5$\cite{ALICE:pOmega}. Compared to a minimum-bias trigger, HM events provide not only a larger number of particles per event, but an overall higher production rate of particles containing strangeness, such a $\Lambda$ hyperons~\cite{ALICE:2017jyt}. Consequently, the HM sample offers a tenfold increase in the amount of \pL pairs reconstructed below \ks of $200~$\MeVc, leading to a total of 1.3 million pairs within the same event sample. The reconstructed primary vertex (PV) of the event is required to have a maximal displacement with respect to the nominal interaction point of 10~cm along the beam axis, in order to ensure a uniform acceptance. Pile-up events with multiple primary vertices are removed following the procedure described in~\cite{ALICE:Run1,ALICE:pXi,ALICE:pOmega,Acharya:2020dfb}. The final number of selected HM events reaches approximately $10^{9}$. Charged particles, such as protons and pions, are directly measured, while the $\Lambda$ candidates are reconstructed based on the IM of the decay products. The correlation functions obtained for particles ($\mathrm{p}\lmb$) and anti-particles ($\overline{\mathrm{p}}\almb$) are identical within uncertainties, thus the final result is presented as their weighted sum $\mathrm{p}\lmb\oplus\overline{\mathrm{p}}\almb$.

Both the protons and the \lmb candidates are reconstructed using the procedure described in~\cite{Acharya:2020dfb}, while the related systematic uncertainties are evaluated by varying the kinematic and topological observables used in the reconstruction. For the purpose of correlation studies it is essential to differentiate between primary particles, which participate in the FSI, and secondary (feed-down) particles, which stem from weak or electromagnetic decays. Experimentally, the former can be selected by demanding the particle candidates to be close to the PV of the event, while the latter have to be associated with a secondary vertex within the event. In the following text, the systematic variations are enclosed in parentheses. The primary proton candidates are selected in the momentum interval $0.5~(0.4,~0.6) < \pt < 4.05$\,\GeVc and $|\eta| < 0.8~(0.77,~0.85)$. To improve the quality of the tracks a minimum of 80 (70, 90) out of the 159 possible spatial points (hits) inside the TPC are required. The candidates are selected by comparing the measurements in the TPC and TOF detectors to the expected distributions for a proton candidate. The agreement is expressed in terms of the detector resolution $\sigma$ ($n^\text{PID}_\sigma$). For protons with $\pt < 0.75$~\GeVc the $n^\text{PID}_\sigma$ is evaluated only based on the energy loss and track measurements in the TPC, while for $\pt > 0.75$~\GeVc a combined TPC and TOF PID selection is applied ($n^\text{PID}_\sigma=\sqrt{n_{\sigma,\mathrm{TPC}}^2+n_{\sigma,\mathrm{TOF}}^2}$). The $n^\text{PID}_\sigma$ of the accepted candidates is required to be within 3 (2.5, 3.5). To reject non-primary particles the distance of closest approach (DCA) to the PV of the tracks is required to be less than 0.1~cm in the transverse plane and less than 0.2~cm along the beam axis. Nevertheless, due to the limited resolution of the reconstruction, the selected primary proton candidates will contain certain amount of secondaries, stemming from weak decays, and misidentifications. These contributions are extracted using Monte Carlo (MC) template fits to the measured distributions of the DCA to the PV~\cite{ALICE:Run1}. The resulting proton purity is $99.4\%$ with a $82.3\%$ fraction of primaries.

The \lmb candidates are reconstructed via the weak decay $\lmb\rightarrow \mathrm{p}\uppi^-$. 
The secondary daughter tracks are subject to similar selection criteria as for the primary protons. In addition, the daughter tracks are required to have a DCA to the PV of at least 0.05~(0.06)~cm. The DCA of the corresponding \lmb candidates to the PV has to be below 1.5~(1.2)~cm. The cosine of the pointing angle (CPA) between the vector connecting the PV to the decay vertex and the three-momentum of the \lmb candidate is required to be larger than 0.99~(0.995). To reject unphysical secondary vertices, reconstructed with tracks stemming from collisions corresponding to different crossings of the beam, the decay tracks are required to possess a hit in one of the SPD or SSD detectors or a matched TOF signal~\cite{ALICE:LL}. The final $\Lambda$ candidates are selected in a $4$~\MeVcc mass window around the nominal mass~\cite{PDG}, where the width of the IM peak is c.a. $1.6~$\MeVcc. The number of primary and secondary contributions for \lmb are extracted similarly as for protons, using the CPA as an observable for the template fits. The average fraction of primary \lmb hyperons is 57.6 (52.1, 60.6)\% and 19.2 (15.4, 21.9)\% originate from the electromagnetic decays of $\Sigma^0$. The number of $\Sigma^0$ particles is related to their ratio to the \lmb hyperons, which is fixed to $0.33~(0.27,~0.40)$. 
These values are based on predictions from the isospin symmetry, thermal model calculations using the Thermal-FIST package~\cite{TFist} and measurements of the corresponding production ratios~\cite{Albrecht:1986me,Sullivan:1987us,Yuldashev:1990az}. Further, each of the weak decays of $\Xi^-$ and $\Xi^0$ contributes with 11.6 (13.5)\% to the yield of \lmb hyperons. The purity of \lmb and \almb was extracted by fitting, as a function of \ks, the IM spectra of candidates selected in the mixed-event sample. The fits were performed in the IM range of 1088 to 1144 \MeVcc using a double Gaussian for the signal and a third-order spline for the background. The result was averaged for $\ks<480$~\MeVc, leading to a purity $P_\lmb=95.3\%$. The systematic variations include a modelling of the signal using the sum of three Gaussians, leading to a purity of $96.3\%$. The effect of misidentified \lmb candidates ($\tilde{\Lambda}$) can be accounted for by the relations
\begin{equation}\label{eq:corr_CF}
    C_\mathrm{exp}(\ks) = P_\lmb C_\mathrm{corrected}(\ks) + (1-P_\lmb)C_{\decpfL},
\end{equation}
\begin{equation}\label{eq:decomp_CF}
    C_\mathrm{corrected}(\ks) = B(\ks) 
    \left[ \lambda_{\decpLpL} C_{\decpLpL}(\ks)+
    \lambda_{\decpLpSo} C_{\decpLpSo}(\ks)+\lambda_{\decpLpXi} C_{\decpLpXi}(\ks)+
    \lambda_{\decpLflat}+\lambda_{\decfpL}
    \right],
\end{equation}
where the signal is decomposed into its ingredients, weighted by the corresponding $\lambda$ parameters and corrected for the non-FSI baseline $B(\ks)$. 

Such a decomposition is required~\cite{ALICE:Run1}, as the experimental signal contains correlations complementing the genuine {\pL} signal $C_{\decpLpL}(\ks)$. In the present analysis the contribution $C_{\decpfL}$ related to misidentified \lmb candidates ($\tilde{\Lambda}$) is explicitly measured and subtracted from the total correlation $C_\mathrm{exp}(\ks)$. This is achieved by performing a sideband analysis~\cite{ALICE:pSig0}, which relies on purposefully selecting $\Lambda$ candidates incompatible with the true $\Lambda$ mass by more than 5$\sigma$.

The corrected correlation $C_\mathrm{corrected}(\ks)$ has an effective $\Lambda$ purity of 100\%, and the remaining contributions (Eq.~\ref{eq:decomp_CF}) are the genuine signal of interest $C_{\decpLpL}$, the residual (feed-down) correlation $C_{\decpLpSo}$ of \lmb particles originating from the decay of a $\Sigma^0$, the residual signal $C_{\decpLpXi}$ related to $\Xi$ ($\Xi^-\oplus\Xi^0$) decaying into \lmb, other sub-dominant (flat) sources of feed-down correlations $C_{\decpLflat}\approx1$, and contamination $C_{\decfpL}$ stemming from misidentified protons. Each of these contributions is weighted by a statistical factor $\lambda$, evaluated as the product of the purities and fractions (primary or secondary) of the set particles~\cite{ALICE:Run1}. These weight factors are summarised in Table~\ref{tab:lambdaval}. 
\begin{table}
\begin{center}
\caption{Weight parameters of the individual components of the \pL correlation function. The two last rows correspond to the values of the $\lambda$ parameters within the systematic variations.}
\begin{tabular}{c|c|c|c|c|c}
\hline \hline
Pair & {\decpLpL} & {\decpLpSo} & {\decpLpXi} & Flat feed-down & {\decfpL} \\
\hline
$\lambda_\mathrm{Pair}$ (\%) & 47.1 & 15.7 & 19.0 & 17.6 & 0.6 \\
min\{$\lambda_\mathrm{Pair}$\} (\%) & 42.7 & 12.6 & -- & -- & -- \\
max\{$\lambda_\mathrm{Pair}$\} (\%) & 49.6 & 18.0 & 22.1 & -- & -- \\
\hline \hline
\end{tabular}
\label{tab:lambdaval}
\end{center}
\end{table} 
The contribution $C_{\decfpL}$ cannot be modelled, however the associated $\lambda_{\decfpL}$ is only 0.6\%, justifying the assumption $\lambda_{\decfpL}C_{\decfpL}\approx\lambda_{\decfpL}$ within the uncertainties of $C_\mathrm{corrected}(\ks)$. By contrast, the residual correlations $C_{\decpLpSo}$ and $C_{\decpLpXi}$ are significant, but in these cases their interactions with protons can be described by theory. Recent correlation studies of the {\pSo} system showed that this interaction is rather weak~\cite{ALICE:pSig0}. This channel is modelled assuming either a flat function or employing the same \Chieft calculations used for the genuine p$\Lambda$ interaction~\cite{Haidenbauer:2019boi}. The contribution from the {\pXi} ($\pXim\oplus\pXio$) channel is modelled employing the lattice potentials from the HAL QCD collaboration~\cite{Sasaki:2019qnh}. They were experimentally validated by comparison with precision measurements of p$\Xi^-$ correlations~\cite{ALICE:pXi,ALICE:pOmega}. The residual contributions $C_{\decpLpSo}(\ks)$ and $C_{\decpLpXi}(\ks)$ are obtained by transforming the corresponding genuine correlation functions to the basis of the p$\Lambda$ interaction, using the formalism described in~\cite{ALICE:Run1} and~\cite{PhysRevC.89.054916} applied to the phase space of the measured pairs.

The non-FSI background (baseline) is parameterised by a third-order polynomial $B(\ks)$ constrained to be flat at $\ks\rightarrow0$ and fitted to the data (Eq.~\ref{eq:decomp_CF}). By default, the fit is performed for $\ks\in[0,456]~$\MeVc, with systematic variations of the upper limit to 432 and 480~\MeVc. Further, due to the expectation of a flat baseline at low \ks, a systematic cross-check has been performed by assuming the hypothesis of a constant $B(\ks)$ and fitting the correlation function for \ks below 336~\MeVc.

The correlation function (Eq.~\ref{eq:decomp_CF}) is given as a function of the measured $k^*$, which is not identical to the true relative momentum of the pair due to the effects of momentum resolution. Thus, to compare the experimental results with theoretical predictions an unfolding of the data is required. Both the same- and mixed-event samples ($N(\ks)$, $M(\ks)$) are biased by the resolution of the detector. They relate to their true underlying distributions by
\begin{equation}\label{eq:ExpNks}
    N(\ks)=\int_0^\infty T(\ks,\kst)N_\mathrm{true}(\kst)d\kst
\end{equation}
and
\begin{equation}\label{eq:ExpMks}
    M(\ks)=\int_0^\infty T(\ks,\kst)M_\mathrm{true}(\kst)d\kst,
\end{equation}
where $T(\ks,\kst)$ is the detector response matrix. The latter is a two-dimensional matrix corresponding to the probability of having a true value \kst given a measured \ks. By using a full scale simulation of the detector, involving Pythia~8~\cite{Sjostrand:2014zea} as an event generator and Geant3~\cite{Brun:1987ma} to model the detector response, the matrix $T(\ks,\kst)$ has been determined. The resulting spread in the distribution of \ks for a fixed \kst is, on average, 4.2~\MeVc. Using $N_\mathrm{true}(\kst)=M_\mathrm{true}(\kst)C(\kst)$ and defining $W(\ks,\kst)=T(\ks,\kst)M_\mathrm{true}(\kst)/M(\ks)$, Eq.~\ref{eq:Cks} becomes equivalent to
\begin{equation}\label{eq:ExpCks}
    C_\text{exp}(\ks)=\EuScript{N}\int_0^\infty W(\ks,\kst)C_\mathrm{true}(\kst)d\kst.
\end{equation}
In the present analysis the unfolding is performed as a two-step process, first obtaining $M_\mathrm{true}(\ks)$ from Eq.~\ref{eq:ExpMks}, second using Eq.~\ref{eq:ExpCks} to obtain $C_\mathrm{true}(\ks)$. Each step is performed by using a cubic spline to parameterise the true functions, which are fitted to their measured counterparts. The splines are defined for $\ks<1000~$\MeVc, using a total of 32 knots. The quality of the procedure is validated by transforming the unfolded functions backwards using Eq.~\ref{eq:ExpMks} and Eq.~\ref{eq:ExpCks}, which ideally should restore the input distributions ($\chi^2=0$). In case the resulting $\chi^2$ per data point is larger than 0.2, the value of each $C_\mathrm{true}(\ks)$ bin is perturbed using a bootstrap procedure~\cite{NumericalRecipes}, until a better $\chi^2$ is achieved. This is iteratively repeated until obtaining the desired precision, and until no single bin deviates by more than half of their uncertainty.

\section{Results and discussion}\label{sec:Results}
\begin{figure}[!ht]
\begin{center}
\includegraphics[width=0.99\textwidth]{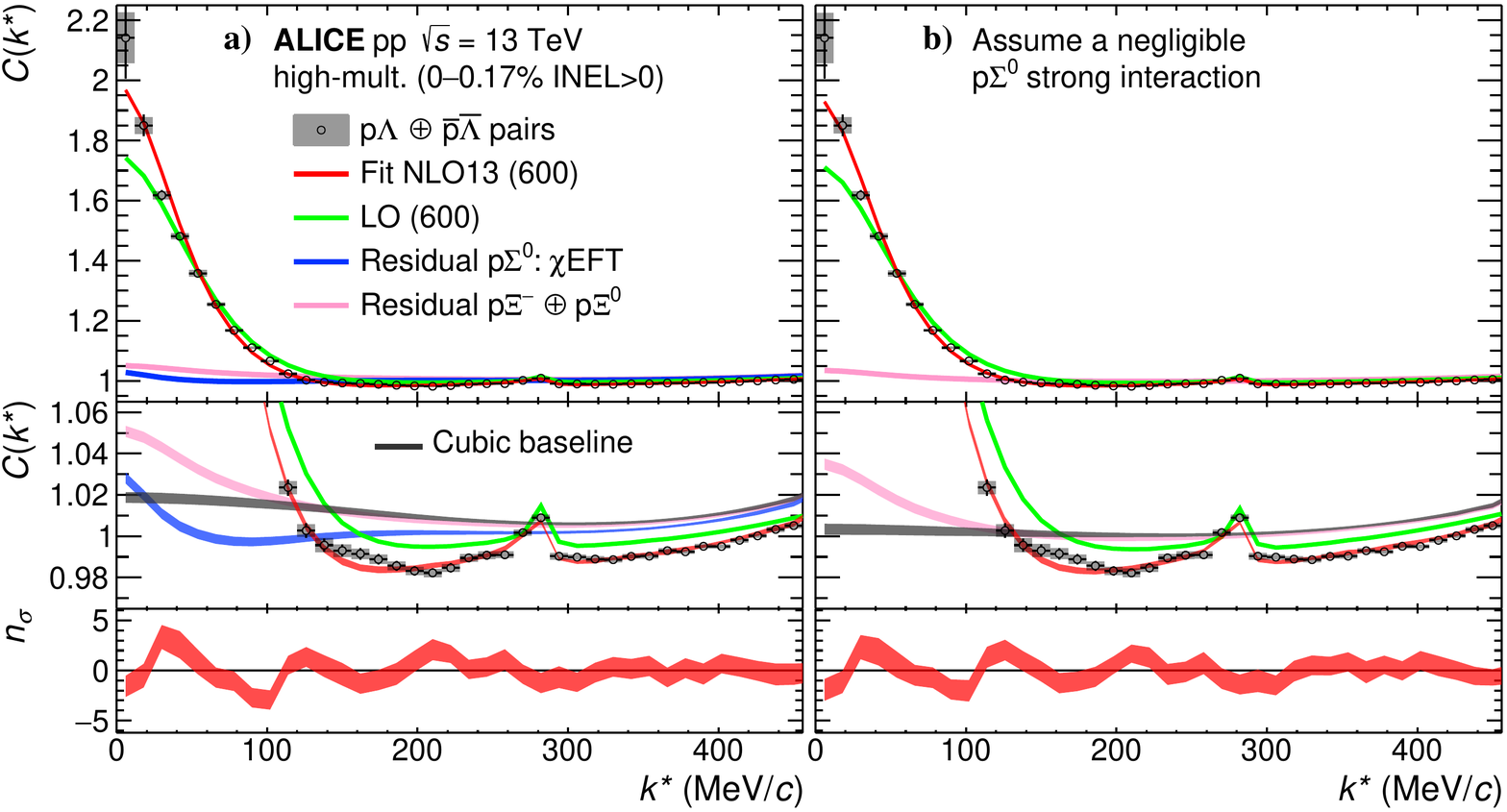}
\end{center}
\caption{
Upper panels: \pL correlation function (circles) with statistical (vertical bars) and systematic (grey boxes) uncertainties. Middle panels: zoom on the cusp-like signal at $\ks = 289$ \MeVc. Lower panels: The deviation between data and predictions, expressed in terms of $n_\sigma$. The fit is performed using NLO13 (red) \Chieft potentials with cut-off $\Lambda=$600~\MeV~\cite{Haidenbauer:2013oca,Haidenbauer:2019boi} and using 
a cubic baseline (dark grey). The residual $\pXim\oplus\pXio$ (pink) and p$\Sigma^0$ (royal blue) correlations are modelled using, respectively, a lattice potential from the HAL QCD collaboration~\cite{Sasaki:2019qnh,ALICE:pXi} and a $\chi$EFT potential~\cite{Haidenbauer:2013oca}. Both contributions are plotted relative to the baseline, while in panel b) the strong interaction of p$\Sigma^0$ is neglected. The reduced $\chi^2$, for $\ks<300~$\MeVc, amounts to 2.2 in case a) and to 1.9 in case b).
}
\label{fig:Results13}
\end{figure}

\begin{figure}[!ht]
\begin{center}
\includegraphics[width=0.99\textwidth]{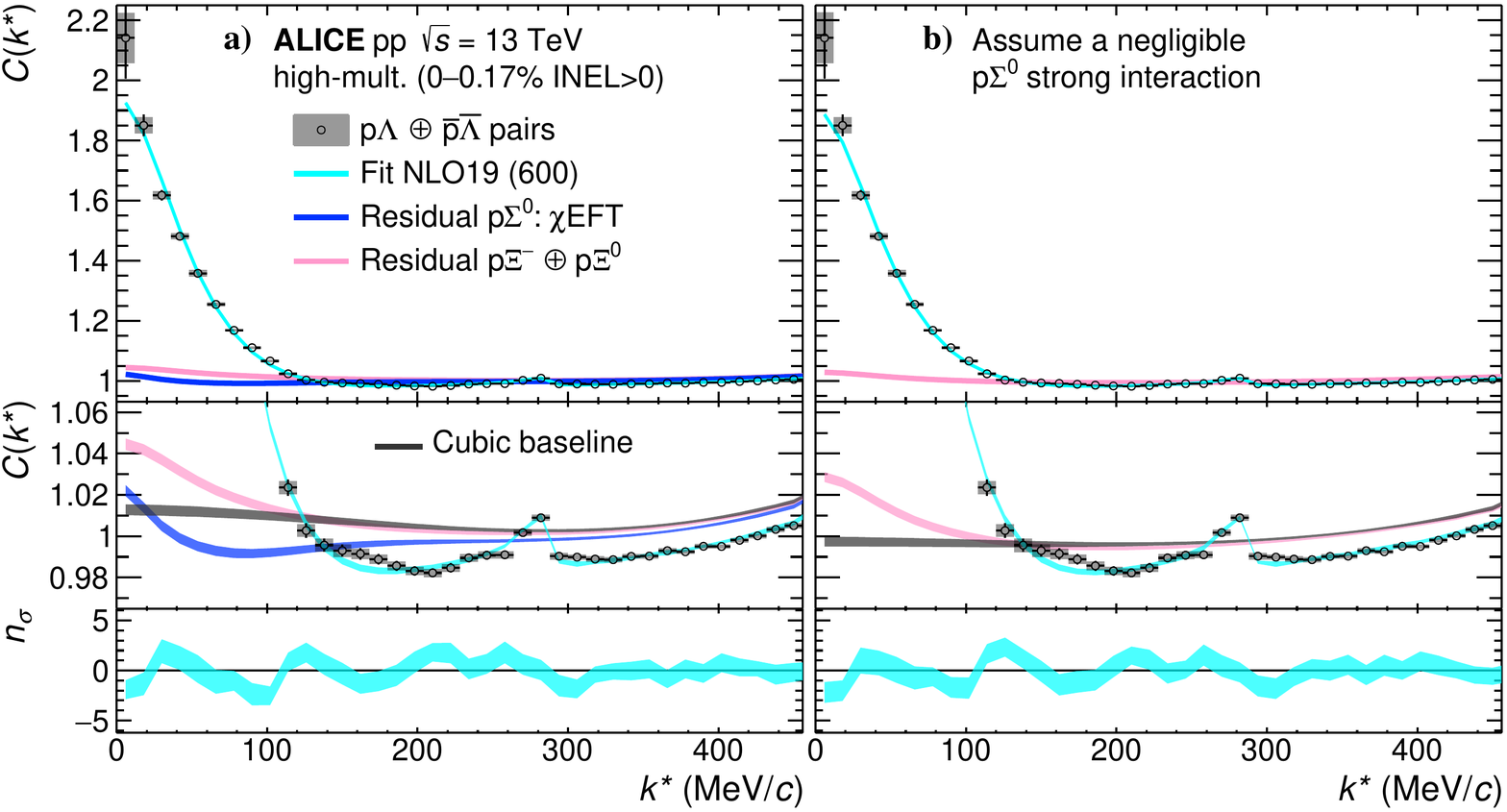}
\end{center}
\caption{
Similar representation as in Fig.~\ref{fig:Results13}, where the \pL interaction is modelled using NLO19 (cyan) \Chieft potentials with cut-off $\Lambda=$600~\MeV~\cite{Haidenbauer:2013oca,Haidenbauer:2019boi}. This leads to an improved description of the low momentum region. The reduced $\chi^2$, for $\ks<300~$\MeVc, equals 2.0 in case the \pSo is modelled by \Chieft (panel a) and 1.8 in case the \pSo final state interaction is ignored (panel b).
}
\label{fig:Results19}
\end{figure}

The corrected and unfolded experimental correlation function for \pL $\oplus$ $\overline{\mathrm{p}}$\almb is shown in Figs.~\ref{fig:Results13} and~\ref{fig:Results19}. The correlation function is measured with high-precision in the low momentum region down to \ks$=6$~\MeVc, in contrast to existing \pL scattering data which cover the region $\ks>$60~\MeVc. The precision achieved for $\ks<$110~\MeVc is better than 1\%, which corresponds to an improvement of factor up to 25 compared to previous scattering data~\cite{SechiZorn:1969hk,Alexander:1969cx,Eisele:1971mk}. The theoretical correlation functions in Eq.~\ref{eq:decomp_CF} were evaluated using the CATS framework~\cite{CATS}. The size of the emitting source employed in the calculation was fixed from independent studies of proton pairs~\cite{Acharya:2020dfb}, which demonstrate a common primordial (core) Gaussian source for pp and p$\Lambda$ pairs when the contribution of strongly decaying resonances is explicitly accounted for~\cite{Acharya:2020dfb}. This source exhibits a pronounced $m_\mathrm{T}$ dependence and considering the average transverse mass $\left<m_\mathrm{T}\right>=1.55~$\GeV of the measured p$\Lambda$ pairs a corresponding core source radius of $r_\mathrm{core}(\left<m_\mathrm{T}\right>)=1.02\pm0.04~$fm is obtained. The total source function can be approximated by an effective Gaussian emission source of size 1.23~fm. The genuine \pL correlation function is modelled by \Chieft hyperon-nucleon potentials, considering the leading-order (LO) interaction~\cite{Polinder:2006zh} and two NLO versions (NLO13~\cite{Haidenbauer:2013oca} and NLO19~\cite{Haidenbauer:2019boi}). 
For the NLO interactions the variation with the underlying cut-off parameter 
$\Lambda$
(cf. Ref.~\cite{Haidenbauer:2013oca}) is explored, while $\Lambda=$600~\MeV is chosen as a default value. Both NLO versions provide an excellent description 
of the available scattering data, having a $\chi^2\approx 16$ for the 
considered 36 data points~\cite{Haidenbauer:2019boi}. 

Figures~\ref{fig:Results13} and~\ref{fig:Results19} show the total fit functions (red and cyan) to the present data. The non-FSI baseline $B(\ks)$ is depicted as a dark grey line, while the individual contributions related to feed-down from $\mathrm{F=\{\Sigma^0,\Xi\}}$ are drawn as royal blue and pink lines, corresponding to $B(\ks)\left[ \lambda_\mathrm{p(F)} C_\mathrm{p(F)}(\ks)+1-\lambda_\mathrm{p(F)}\right]$. The latter relation is derived by setting all $C_{i}$ terms within Eq.~\ref{eq:decomp_CF}, apart from $C_\mathrm{p(F)}$, equal to unity. The upper panels in Figs.~\ref{fig:Results13} and~\ref{fig:Results19} present the correlation function in the whole \ks range, while the middle panels show the region where the \SN channels open, clearly visible as a cusp structure occurring at $\ks=289$~\MeVc. The deviation between data and prediction, expressed in terms of number of standard deviations $n_\sigma$, is shown in the bottom panels. The discrepancy between theory and data is largest in the momentum region $\ks<$110~\MeVc, while, due to the presence of the N$\Sigma$ cusp, the sensitivity of the correlation function to the properties of the strong interaction extends up to 300~\MeVc. The deviations for the interaction hypotheses are summarised in Table~\ref{tab:dsig_plb}, where the left two columns show the $n_\sigma$ only in the low momentum region, and the right two columns represent the deviation evaluated for $\ks\in[0,300]$~\MeVc.

The presented results are the first direct experimental evidence of the \SNLN coupling in a two-body final state. The signal of the cusp is determined by the properties of the interaction, and further modified by the relative amount of N$\Sigma$ and p$\Lambda$ initial state pairs leading to the final state (measured) p$\Lambda$ pairs. The amount of initial state pairs was fixed by the above-mentioned $\Sigma$:$\Lambda$ ratio, enabling a direct test of the strong interaction.
The LO chiral potential~\cite{Polinder:2006zh} predicts a too small \SN cusp 
with respect to the measurement, the green line in Fig.~\ref{fig:Results13}, while both 
NLO interactions provide a satisfactory description of the cusp structure. 
On the other hand, in the momentum region below 110~\MeVc 
there is a tension between the data and the theory predictions for 
all considered interactions. In particular, the results for the two 
NLO potentials are not that well in line with the measured correlation function,
despite of the fact that these interactions reproduce the 
low-energy p$\Lambda$ scattering data perfectly~\cite{Haidenbauer:2019boi}. 
The best result is provided by the NLO19 potential with  
$\Lambda=$600--650~\MeV, though the deviation of $n_\sigma=3.2$ from the experiment is substantial. 
For NLO13 this deviation is even larger and amounts to $n_\sigma=4.2$. Further, it is observed that for NLO13 and NLO19 the best agreement with the data is achieved within the same range of cut-off values (550--650~\MeV) which also provide the best description of the available scattering and hypertriton data~\cite{Haidenbauer:2013oca,Haidenbauer:2019boi}. 

The discrepancy between the data and \Chieft at low momenta could be an 
indication for a weaker genuine \pL interaction, but it could also 
signal that the \pSo correlation is very small. As visible in the right panels 
of Figs.~\ref{fig:Results13},~\ref{fig:Results19} and Table~\ref{tab:dsig_plb}, adopting the hypothesis of a negligible \pSo correlation leads to a 
better agreement with the present \pL data ($n_\sigma=2.2$).
At the moment it is impossible to differentiate between these two cases 
because the existing direct measurement of the \pSo channel is not precise enough for drawing pertinent conclusions~\cite{ALICE:pSig0}. The \pSo measurement is compatible with both the NLO predictions (of a weakly attractive p$\Sigma^0$ interaction) and with a flat correlation (negligible p$\Sigma^0$ interaction). A precision measurement of the genuine \pSo channel, expected to be achieved in the upcoming LHC Run 3~\cite{ALICE-Run3}, should provide clarification. Then the actual strength of the N$\Lambda$ interaction can be pinned down in a model independent way by a dedicated theoretical analysis of the p$\Lambda$ data.  

All the conclusions of the present analysis remain the same under the alternative hypothesis of a constant baseline, or in case the deviation is evaluated for $\ks<300~$\MeVc. Within that momentum region, the NLO19 provides a satisfactory description of the data, with a deviation of $n_\sigma=2.8$, while the NLO13 still results in a larger discrepancy ($n_\sigma=3.6$).

\begin{table} 
\begin{center} 
\caption{The deviation, expressed in terms of $n_\sigma$, between data and prediction for the different interaction hypotheses of $\text{p}\Lambda$ and $\text{p}\Sigma^0$, evaluated for $k^{*}\in[0,110]~\MeVc$ (first two columns) and $k^{*}\in[0,300]~\MeVc$ (last two columns). The default values correspond to the fit with a cubic baseline and the values in parentheses represent the results using a constant baseline. The default interaction (in bold) is the $\chi$EFT NLO19 potential with cut-off $\Lambda=$600~\MeV~\cite{Haidenbauer:2019boi}. 
Each row corresponds to a different variant of the $\chi$EFT interaction 
used for evaluating the \pL correlation. The first and third column correspond to the case of modelling the $\text{p}\Sigma^0$ using \Chieft, while the second and fourth column represent the case of negligible $\text{p}\Sigma^0$ final state interaction.}
\begin{tabular}{r||c|c||c|c} 
\hline \hline 
 & \multicolumn{4}{c}{Standard deviation ($n_\sigma$)} \\
\cline{2-5}
 & \multicolumn{2}{c||}{$\ks\in[0,110~]$\MeVc} & \multicolumn{2}{c}{$\ks\in[0,300~]$\MeVc} \\
 \cline{2-5}
$\text{p}\Sigma^0$ ($\rightarrow$)& \pmb{$\chi$}\textbf{EFT} & Negligible & \pmb{$\chi$}\textbf{EFT} & Negligible \\ 
$\text{p}\Lambda$ ($\downarrow$) & & p$\Sigma^0$ FSI & & p$\Sigma^0$ FSI \\ 
\hline 
LO-600 & 4.7 (4.9) & 6.1 (7.0) & 7.2 (8.7) & 10.3 (10.3) \\ 
NLO13-500 & 5.9 (8.0) & 4.3 (5.1) & 6.6 (10.3) & 4.9 (7.6)\\ 
NLO13-550 & 4.5 (5.8) & 3.1 (3.1) & 4.1 (7.2) & 2.8 (3.4)\\ 
NLO13-600 & 4.5 (5.3) & 3.2 (3.1) & 3.9 (5.1) & 2.9 (3.0)\\ 
NLO13-650 & 4.2 (4.7) & 2.8 (2.7) & 3.6 (4.1) & 2.8 (3.3)\\ 
NLO19-500 & 4.2 (5.0) & 2.7 (3.0) & 4.4 (7.6) & 3.4 (4.3)\\ 
NLO19-550 & 3.6 (4.2) & 2.4 (2.7) & 3.0 (4.4) & 2.2 (2.7)\\ 
\textbf{NLO19-600} & 3.2 (3.2) & 2.2 (2.3) & 3.1 (3.8) & 2.6 (3.3)\\ 
NLO19-650 & 3.2 (3.6) & 2.3 (2.0) & 2.8 (3.2) & 2.7 (3.5)\\ 
\hline \hline 
\end{tabular} 
\label{tab:dsig_plb} 
\end{center} 
\end{table}

\section{Summary}\label{sec:Summary}

In conclusion, two-particle correlation techniques were used to study the final state interaction in the \SNLN coupled system. This was achieved by studying the \pL correlation function at low relative momenta with an unprecedented precision. The significance of the coupling of p$\Lambda$ to N$\Sigma$ is manifested as a cusp-like enhancement present at the corresponding threshold energy, which is the first direct experimental observation of this structure. Further, using different modellings for the \pSo feed-down leads to a statistically significant modification of the measured \pL correlation, implying an indirect sensitivity to the genuine \pSo correlation. 
In the momentum range $\ks\in[110,300]$~\MeVc all of the tested NLO \Chieft interactions are compatible with the data, however a significant deviation is present at lower values. The detailed analysis, presented in Table~\ref{tab:dsig_plb}, reveals a 
deviation of at least $n_\sigma=3.2$, for $\ks<110~$\MeVc, for 
the considered \Chieft interactions. The result for NLO19 exhibits an overall better compatibility, compared to the NLO13 prediction. The former involves a weaker \SNLN transition potential and a more attractive two-body interaction of the \lmb hyperon in the medium. This requires a stronger repulsive NN$\Lambda$ three-body force, which leads to a stiffening of the EoS at large densities~\cite{Gerstung:2020ktv} and a disfavoured production of these strange hadrons in neutron stars. The presented data provide an opportunity to improve the theoretical calculations for the \SNLN coupled system, including the low-energy properties of N$\Lambda$. 
The successful use of correlation techniques in the two-body sector can be extended to measure directly the three-body correlations~\cite{DelGrande:2021mju}.
The increased amount of statistics during the third running period of the LHC~\cite{ALICE-Run3} will allow for such measurements.

%%%%% acknowledgements
\newenvironment{acknowledgement}{\relax}{\relax}
\begin{acknowledgement}
\section*{Acknowledgements}
% Version: 2021-03-10

The ALICE Collaboration would like to thank all its engineers and technicians for their invaluable contributions to the construction of the experiment and the CERN accelerator teams for the outstanding performance of the LHC complex.
The ALICE Collaboration gratefully acknowledges the resources and support provided by all Grid centres and the Worldwide LHC Computing Grid (WLCG) collaboration.
The ALICE Collaboration acknowledges the following funding agencies for their support in building and running the ALICE detector:
A. I. Alikhanyan National Science Laboratory (Yerevan Physics Institute) Foundation (ANSL), State Committee of Science and World Federation of Scientists (WFS), Armenia;
Austrian Academy of Sciences, Austrian Science Fund (FWF): [M 2467-N36] and Nationalstiftung f\"{u}r Forschung, Technologie und Entwicklung, Austria;
Ministry of Communications and High Technologies, National Nuclear Research Center, Azerbaijan;
Conselho Nacional de Desenvolvimento Cient\'{\i}fico e Tecnol\'{o}gico (CNPq), Financiadora de Estudos e Projetos (Finep), Funda\c{c}\~{a}o de Amparo \`{a} Pesquisa do Estado de S\~{a}o Paulo (FAPESP) and Universidade Federal do Rio Grande do Sul (UFRGS), Brazil;
Ministry of Education of China (MOEC) , Ministry of Science \& Technology of China (MSTC) and National Natural Science Foundation of China (NSFC), China;
Ministry of Science and Education and Croatian Science Foundation, Croatia;
Centro de Aplicaciones Tecnol\'{o}gicas y Desarrollo Nuclear (CEADEN), Cubaenerg\'{\i}a, Cuba;
Ministry of Education, Youth and Sports of the Czech Republic, Czech Republic;
The Danish Council for Independent Research | Natural Sciences, the VILLUM FONDEN and Danish National Research Foundation (DNRF), Denmark;
Helsinki Institute of Physics (HIP), Finland;
Commissariat \`{a} l'Energie Atomique (CEA) and Institut National de Physique Nucl\'{e}aire et de Physique des Particules (IN2P3) and Centre National de la Recherche Scientifique (CNRS), France;
Bundesministerium f\"{u}r Bildung und Forschung (BMBF) and GSI Helmholtzzentrum f\"{u}r Schwerionenforschung GmbH, Germany;
General Secretariat for Research and Technology, Ministry of Education, Research and Religions, Greece;
National Research, Development and Innovation Office, Hungary;
Department of Atomic Energy Government of India (DAE), Department of Science and Technology, Government of India (DST), University Grants Commission, Government of India (UGC) and Council of Scientific and Industrial Research (CSIR), India;
Indonesian Institute of Science, Indonesia;
Istituto Nazionale di Fisica Nucleare (INFN), Italy;
Institute for Innovative Science and Technology , Nagasaki Institute of Applied Science (IIST), Japanese Ministry of Education, Culture, Sports, Science and Technology (MEXT) and Japan Society for the Promotion of Science (JSPS) KAKENHI, Japan;
Consejo Nacional de Ciencia (CONACYT) y Tecnolog\'{i}a, through Fondo de Cooperaci\'{o}n Internacional en Ciencia y Tecnolog\'{i}a (FONCICYT) and Direcci\'{o}n General de Asuntos del Personal Academico (DGAPA), Mexico;
Nederlandse Organisatie voor Wetenschappelijk Onderzoek (NWO), Netherlands;
The Research Council of Norway, Norway;
Commission on Science and Technology for Sustainable Development in the South (COMSATS), Pakistan;
Pontificia Universidad Cat\'{o}lica del Per\'{u}, Peru;
Ministry of Education and Science, National Science Centre and WUT ID-UB, Poland;
Korea Institute of Science and Technology Information and National Research Foundation of Korea (NRF), Republic of Korea;
Ministry of Education and Scientific Research, Institute of Atomic Physics and Ministry of Research and Innovation and Institute of Atomic Physics, Romania;
Joint Institute for Nuclear Research (JINR), Ministry of Education and Science of the Russian Federation, National Research Centre Kurchatov Institute, Russian Science Foundation and Russian Foundation for Basic Research, Russia;
Ministry of Education, Science, Research and Sport of the Slovak Republic, Slovakia;
National Research Foundation of South Africa, South Africa;
Swedish Research Council (VR) and Knut \& Alice Wallenberg Foundation (KAW), Sweden;
European Organization for Nuclear Research, Switzerland;
Suranaree University of Technology (SUT), National Science and Technology Development Agency (NSDTA) and Office of the Higher Education Commission under NRU project of Thailand, Thailand;
Turkish Atomic Energy Agency (TAEK), Turkey;
National Academy of  Sciences of Ukraine, Ukraine;
Science and Technology Facilities Council (STFC), United Kingdom;
National Science Foundation of the United States of America (NSF) and United States Department of Energy, Office of Nuclear Physics (DOE NP), United States of America.   
\end{acknowledgement}

%%%%%%%% Bibliography (In case of using bibtex generate the bbl requested by arXiv)
%\bibliographystyle{utphys}   % Remember we use title in the biblio
%\bibliography{biblio_intro}
%\bibliography{bibliog.bib}
%\input {bibliography.tex}  

\bibliographystyle{utphys}
%\bibliographystyle{plain}
%\nocite{*} %--> to check if there are any uncited ref
\bibliography{bibliog.bib}

%%%%%%%%% appendix with author list
\newpage
\appendix

\section{The ALICE Collaboration}
\label{app:collab}
% Collaboration: CERN-LHC-ALICE
% Generation Date is 2020-05-17

% How to use:
%%%%%%%%% appendix with author list
%\appendix
%\section{The ALICE Collaboration}
%\label{app:collab}
%\input{Alice_Authorslist_XXXX-Axx-XX.tex}
\begingroup
\small
\begin{flushleft}
S.~Acharya\Irefn{org142}\And 
D.~Adamov\'{a}\Irefn{org97}\And 
A.~Adler\Irefn{org75}\And 
J.~Adolfsson\Irefn{org82}\And 
G.~Aglieri Rinella\Irefn{org35}\And 
M.~Agnello\Irefn{org31}\And 
N.~Agrawal\Irefn{org55}\And 
Z.~Ahammed\Irefn{org142}\And 
S.~Ahmad\Irefn{org16}\And 
S.U.~Ahn\Irefn{org77}
% ...
\And
\renewcommand\labelenumi{\textsuperscript{\theenumi}~}

\section*{Affiliation notes}
\renewcommand\theenumi{\roman{enumi}}
\begin{Authlist}
\item \Adef{orgI}Deceased
\item \Adef{orgII}Italian National Agency for New Technologies, Energy and Sustainable Economic Development (ENEA), Bologna, Italy
\item \Adef{orgIII}Dipartimento DET del Politecnico di Torino, Turin, Italy
\item \Adef{orgIV}M.V. Lomonosov Moscow State University, D.V. Skobeltsyn Institute of Nuclear, Physics, Moscow, Russia
\item \Adef{orgV}Institute of Theoretical Physics, University of Wroclaw, Poland
\end{Authlist}

\section*{Collaboration Institutes}
\renewcommand\theenumi{\arabic{enumi}~}
\begin{Authlist}
\item \Idef{org1} A.I. Alikhanyan National Science Laboratory (Yerevan Physics Institute) Foundation, Yerevan, Armenia
\item \Idef{org2} AGH University of Science and Technology, Cracow, Poland
\item \Idef{org3} Bogolyubov Institute for Theoretical Physics, National Academy of Sciences of Ukraine, Kiev, Ukraine
\item \Idef{org4} Bose Institute, Department of Physics  and Centre for Astroparticle Physics and Space Science (CAPSS), Kolkata, India
\item \Idef{org5} Budker Institute for Nuclear Physics, Novosibirsk, Russia
\item \Idef{org6} California Polytechnic State University, San Luis Obispo, California, United States
\item \Idef{org7} Central China Normal University, Wuhan, China
\item \Idef{org8} Centro de Aplicaciones Tecnol\'{o}gicas y Desarrollo Nuclear (CEADEN), Havana, Cuba
\item \Idef{org9} Centro de Investigaci\'{o}n y de Estudios Avanzados (CINVESTAV), Mexico City and M\'{e}rida, Mexico
\item \Idef{org10} Chicago State University, Chicago, Illinois, United States
\item \Idef{org11} China Institute of Atomic Energy, Beijing, China
\item \Idef{org12} Chungbuk National University, Cheongju, Republic of Korea
\item \Idef{org13} Comenius University Bratislava, Faculty of Mathematics, Physics and Informatics, Bratislava, Slovakia
\item \Idef{org14} COMSATS University Islamabad, Islamabad, Pakistan
\item \Idef{org15} Creighton University, Omaha, Nebraska, United States
\item \Idef{org16} Department of Physics, Aligarh Muslim University, Aligarh, India
\item \Idef{org17} Department of Physics, Pusan National University, Pusan, Republic of Korea
\item \Idef{org18} Department of Physics, Sejong University, Seoul, Republic of Korea
\item \Idef{org19} Department of Physics, University of California, Berkeley, California, United States
\item \Idef{org20} Department of Physics, University of Oslo, Oslo, Norway
\item \Idef{org21} Department of Physics and Technology, University of Bergen, Bergen, Norway
\item \Idef{org22} Dipartimento di Fisica dell'Universit\`{a} 'La Sapienza' and Sezione INFN, Rome, Italy
\item \Idef{org23} Dipartimento di Fisica dell'Universit\`{a} and Sezione INFN, Cagliari, Italy
\item \Idef{org24} Dipartimento di Fisica dell'Universit\`{a} and Sezione INFN, Trieste, Italy
\item \Idef{org25} Dipartimento di Fisica dell'Universit\`{a} and Sezione INFN, Turin, Italy
\item \Idef{org26} Dipartimento di Fisica e Astronomia dell'Universit\`{a} and Sezione INFN, Bologna, Italy
\item \Idef{org27} Dipartimento di Fisica e Astronomia dell'Universit\`{a} and Sezione INFN, Catania, Italy
\item \Idef{org28} Dipartimento di Fisica e Astronomia dell'Universit\`{a} and Sezione INFN, Padova, Italy
\item \Idef{org29} Dipartimento di Fisica e Nucleare e Teorica, Universit\`{a} di Pavia, Pavia, Italy
\item \Idef{org30} Dipartimento di Fisica `E.R.~Caianiello' dell'Universit\`{a} and Gruppo Collegato INFN, Salerno, Italy
\item \Idef{org31} Dipartimento DISAT del Politecnico and Sezione INFN, Turin, Italy
\item \Idef{org32} Dipartimento di Scienze e Innovazione Tecnologica dell'Universit\`{a} del Piemonte Orientale and INFN Sezione di Torino, Alessandria, Italy
\item \Idef{org33} Dipartimento di Scienze MIFT, Universit\`{a} di Messina, Messina, Italy
\item \Idef{org34} Dipartimento Interateneo di Fisica `M.~Merlin' and Sezione INFN, Bari, Italy
\item \Idef{org35} European Organization for Nuclear Research (CERN), Geneva, Switzerland
\item \Idef{org36} Faculty of Electrical Engineering, Mechanical Engineering and Naval Architecture, University of Split, Split, Croatia
\item \Idef{org37} Faculty of Engineering and Science, Western Norway University of Applied Sciences, Bergen, Norway
\item \Idef{org38} Faculty of Nuclear Sciences and Physical Engineering, Czech Technical University in Prague, Prague, Czech Republic
\item \Idef{org39} Faculty of Science, P.J.~\v{S}af\'{a}rik University, Ko\v{s}ice, Slovakia
\item \Idef{org40} Frankfurt Institute for Advanced Studies, Johann Wolfgang Goethe-Universit\"{a}t Frankfurt, Frankfurt, Germany
\item \Idef{org41} Fudan University, Shanghai, China
\item \Idef{org42} Gangneung-Wonju National University, Gangneung, Republic of Korea
\item \Idef{org43} Gauhati University, Department of Physics, Guwahati, India
\item \Idef{org44} Helmholtz-Institut f\"{u}r Strahlen- und Kernphysik, Rheinische Friedrich-Wilhelms-Universit\"{a}t Bonn, Bonn, Germany
\item \Idef{org45} Helsinki Institute of Physics (HIP), Helsinki, Finland
\item \Idef{org46} High Energy Physics Group,  Universidad Aut\'{o}noma de Puebla, Puebla, Mexico
\item \Idef{org47} Hiroshima University, Hiroshima, Japan
\item \Idef{org48} Hochschule Worms, Zentrum  f\"{u}r Technologietransfer und Telekommunikation (ZTT), Worms, Germany
\item \Idef{org49} Horia Hulubei National Institute of Physics and Nuclear Engineering, Bucharest, Romania
\item \Idef{org50} Indian Institute of Technology Bombay (IIT), Mumbai, India
\item \Idef{org51} Indian Institute of Technology Indore, Indore, India
\item \Idef{org52} Indonesian Institute of Sciences, Jakarta, Indonesia
\item \Idef{org53} INFN, Laboratori Nazionali di Frascati, Frascati, Italy
\item \Idef{org54} INFN, Sezione di Bari, Bari, Italy
\item \Idef{org55} INFN, Sezione di Bologna, Bologna, Italy
\item \Idef{org56} INFN, Sezione di Cagliari, Cagliari, Italy
\item \Idef{org57} INFN, Sezione di Catania, Catania, Italy
\item \Idef{org58} INFN, Sezione di Padova, Padova, Italy
\item \Idef{org59} INFN, Sezione di Roma, Rome, Italy
\item \Idef{org60} INFN, Sezione di Torino, Turin, Italy
\item \Idef{org61} INFN, Sezione di Trieste, Trieste, Italy
\item \Idef{org62} Inha University, Incheon, Republic of Korea
\item \Idef{org149} Institute for Advanced Simulation, Forschungszentrum J\"ulich, J\"ulich, Germany
\item \Idef{org63} Institute for Gravitational and Subatomic Physics (GRASP), Utrecht University/Nikhef, Utrecht, Netherlands
\item \Idef{org64} Institute for Nuclear Research, Academy of Sciences, Moscow, Russia
\item \Idef{org65} Institute of Experimental Physics, Slovak Academy of Sciences, Ko\v{s}ice, Slovakia
\item \Idef{org66} Institute of Physics, Homi Bhabha National Institute, Bhubaneswar, India
\item \Idef{org67} Institute of Physics of the Czech Academy of Sciences, Prague, Czech Republic
\item \Idef{org68} Institute of Space Science (ISS), Bucharest, Romania
\item \Idef{org69} Institut f\"{u}r Kernphysik, Johann Wolfgang Goethe-Universit\"{a}t Frankfurt, Frankfurt, Germany
\item \Idef{org70} Instituto de Ciencias Nucleares, Universidad Nacional Aut\'{o}noma de M\'{e}xico, Mexico City, Mexico
\item \Idef{org71} Instituto de F\'{i}sica, Universidade Federal do Rio Grande do Sul (UFRGS), Porto Alegre, Brazil
\item \Idef{org72} Instituto de F\'{\i}sica, Universidad Nacional Aut\'{o}noma de M\'{e}xico, Mexico City, Mexico
\item \Idef{org73} iThemba LABS, National Research Foundation, Somerset West, South Africa
\item \Idef{org74} Jeonbuk National University, Jeonju, Republic of Korea
\item \Idef{org75} Johann-Wolfgang-Goethe Universit\"{a}t Frankfurt Institut f\"{u}r Informatik, Fachbereich Informatik und Mathematik, Frankfurt, Germany
\item \Idef{org76} Joint Institute for Nuclear Research (JINR), Dubna, Russia
\item \Idef{org77} Korea Institute of Science and Technology Information, Daejeon, Republic of Korea
\item \Idef{org78} KTO Karatay University, Konya, Turkey
\item \Idef{org79} Laboratoire de Physique des 2 Infinis, Ir\`{e}ne Joliot-Curie, Orsay, France
\item \Idef{org80} Laboratoire de Physique Subatomique et de Cosmologie, Universit\'{e} Grenoble-Alpes, CNRS-IN2P3, Grenoble, France
\item \Idef{org81} Lawrence Berkeley National Laboratory, Berkeley, California, United States
\item \Idef{org82} Lund University Department of Physics, Division of Particle Physics, Lund, Sweden
\item \Idef{org83} Moscow Institute for Physics and Technology, Moscow, Russia
\item \Idef{org84} Nagasaki Institute of Applied Science, Nagasaki, Japan
\item \Idef{org85} Nara Women{'}s University (NWU), Nara, Japan
\item \Idef{org86} National and Kapodistrian University of Athens, School of Science, Department of Physics , Athens, Greece
\item \Idef{org87} National Centre for Nuclear Research, Warsaw, Poland
\item \Idef{org88} National Institute of Science Education and Research, Homi Bhabha National Institute, Jatni, India
\item \Idef{org89} National Nuclear Research Center, Baku, Azerbaijan
\item \Idef{org90} National Research Centre Kurchatov Institute, Moscow, Russia
\item \Idef{org91} Niels Bohr Institute, University of Copenhagen, Copenhagen, Denmark
\item \Idef{org92} Nikhef, National institute for subatomic physics, Amsterdam, Netherlands
\item \Idef{org93} NRC Kurchatov Institute IHEP, Protvino, Russia
\item \Idef{org94} NRC \guillemotleft Kurchatov\guillemotright  Institute - ITEP, Moscow, Russia
\item \Idef{org95} NRNU Moscow Engineering Physics Institute, Moscow, Russia
\item \Idef{org96} Nuclear Physics Group, STFC Daresbury Laboratory, Daresbury, United Kingdom
\item \Idef{org97} Nuclear Physics Institute of the Czech Academy of Sciences, \v{R}e\v{z} u Prahy, Czech Republic
\item \Idef{org98} Oak Ridge National Laboratory, Oak Ridge, Tennessee, United States
\item \Idef{org99} Ohio State University, Columbus, Ohio, United States
\item \Idef{org100} Petersburg Nuclear Physics Institute, Gatchina, Russia
\item \Idef{org101} Physics department, Faculty of science, University of Zagreb, Zagreb, Croatia
\item \Idef{org102} Physics Department, Panjab University, Chandigarh, India
\item \Idef{org103} Physics Department, University of Jammu, Jammu, India
\item \Idef{org104} Physics Department, University of Rajasthan, Jaipur, India
\item \Idef{org105} Physikalisches Institut, Eberhard-Karls-Universit\"{a}t T\"{u}bingen, T\"{u}bingen, Germany
\item \Idef{org106} Physikalisches Institut, Ruprecht-Karls-Universit\"{a}t Heidelberg, Heidelberg, Germany
\item \Idef{org107} Physik Department, Technische Universit\"{a}t M\"{u}nchen, Munich, Germany
\item \Idef{org108} Politecnico di Bari and Sezione INFN, Bari, Italy
\item \Idef{org109} Research Division and ExtreMe Matter Institute EMMI, GSI Helmholtzzentrum f\"ur Schwerionenforschung GmbH, Darmstadt, Germany
\item \Idef{org110} Russian Federal Nuclear Center (VNIIEF), Sarov, Russia
\item \Idef{org111} Saha Institute of Nuclear Physics, Homi Bhabha National Institute, Kolkata, India
\item \Idef{org112} School of Physics and Astronomy, University of Birmingham, Birmingham, United Kingdom
\item \Idef{org113} Secci\'{o}n F\'{\i}sica, Departamento de Ciencias, Pontificia Universidad Cat\'{o}lica del Per\'{u}, Lima, Peru
\item \Idef{org114} St. Petersburg State University, St. Petersburg, Russia
\item \Idef{org115} Stefan Meyer Institut f\"{u}r Subatomare Physik (SMI), Vienna, Austria
\item \Idef{org116} SUBATECH, IMT Atlantique, Universit\'{e} de Nantes, CNRS-IN2P3, Nantes, France
\item \Idef{org117} Suranaree University of Technology, Nakhon Ratchasima, Thailand
\item \Idef{org118} Technical University of Ko\v{s}ice, Ko\v{s}ice, Slovakia
\item \Idef{org119} The Henryk Niewodniczanski Institute of Nuclear Physics, Polish Academy of Sciences, Cracow, Poland
\item \Idef{org120} The University of Texas at Austin, Austin, Texas, United States
\item \Idef{org121} Universidad Aut\'{o}noma de Sinaloa, Culiac\'{a}n, Mexico
\item \Idef{org122} Universidade de S\~{a}o Paulo (USP), S\~{a}o Paulo, Brazil
\item \Idef{org123} Universidade Estadual de Campinas (UNICAMP), Campinas, Brazil
\item \Idef{org124} Universidade Federal do ABC, Santo Andre, Brazil
\item \Idef{org125} University of Cape Town, Cape Town, South Africa
\item \Idef{org126} University of Houston, Houston, Texas, United States
\item \Idef{org127} University of Jyv\"{a}skyl\"{a}, Jyv\"{a}skyl\"{a}, Finland
\item \Idef{org128} University of Kansas, Lawrence, Kansas, United States
\item \Idef{org129} University of Liverpool, Liverpool, United Kingdom
\item \Idef{org130} University of Science and Technology of China, Hefei, China
\item \Idef{org131} University of South-Eastern Norway, Tonsberg, Norway
\item \Idef{org132} University of Tennessee, Knoxville, Tennessee, United States
\item \Idef{org133} University of the Witwatersrand, Johannesburg, South Africa
\item \Idef{org134} University of Tokyo, Tokyo, Japan
\item \Idef{org135} University of Tsukuba, Tsukuba, Japan
\item \Idef{org136} Universit\'{e} Clermont Auvergne, CNRS/IN2P3, LPC, Clermont-Ferrand, France
\item \Idef{org137} Universit\'{e} de Lyon, CNRS/IN2P3, Institut de Physique des 2 Infinis de Lyon , Lyon, France
\item \Idef{org138} Universit\'{e} de Strasbourg, CNRS, IPHC UMR 7178, F-67000 Strasbourg, France, Strasbourg, France
\item \Idef{org139} Universit\'{e} Paris-Saclay Centre d'Etudes de Saclay (CEA), IRFU, D\'{e}partment de Physique Nucl\'{e}aire (DPhN), Saclay, France
\item \Idef{org140} Universit\`{a} degli Studi di Foggia, Foggia, Italy
\item \Idef{org141} Universit\`{a} di Brescia, Brescia, Italy
\item \Idef{org142} Variable Energy Cyclotron Centre, Homi Bhabha National Institute, Kolkata, India
\item \Idef{org143} Warsaw University of Technology, Warsaw, Poland
\item \Idef{org144} Wayne State University, Detroit, Michigan, United States
\item \Idef{org145} Westf\"{a}lische Wilhelms-Universit\"{a}t M\"{u}nster, Institut f\"{u}r Kernphysik, M\"{u}nster, Germany
\item \Idef{org146} Wigner Research Centre for Physics, Budapest, Hungary
\item \Idef{org147} Yale University, New Haven, Connecticut, United States
\item \Idef{org148} Yonsei University, Seoul, Republic of Korea
\end{Authlist}
\endgroup  %%%%%%% done by webmaster team
\end{document}